\documentclass[prd, amsfonts, twocolumn, nofootinbib, showpacs]{revtex4-2}
\usepackage{graphicx, epsfig}
\usepackage{color}
\usepackage{amsmath}
\usepackage{amssymb}
\usepackage{physics}
\newcommand{\be}{\begin{equation}}
\newcommand{\ee}{\end{equation}}
\newcommand{\bea}{\begin{eqnarray}}
\newcommand{\eea}{\end{eqnarray}}

\newcommand{\gapp}{\mathrel{\raise.3ex\hbox{$>$}\mkern-14mu
\lower0.6ex\hbox{$\sim$}}}
\newcommand{\lapp}{\mathrel{\raise.3ex\hbox{$<$}\mkern-14mu
\lower0.6ex\hbox{$\sim$}}}
\def\bbox{{\,\lower0.9pt\vbox{\hrule \hbox{\vrule height 0.2 cm
\hskip 0.2 cm \vrule  height 0.2 cm}\hrule}\,}}

\begin{document}
\title{Shedding new light on the absence of fermionic superradiance and maximal infalling rate of fermions into a black hole 
}
\author{De-Chang Dai$^{1,2}$\footnote{communicating author: De-Chang Dai,\\ email: diedachung@gmail.com\label{fnlabel}}}
\affiliation{ $^1$ Department of Physics, National Dong Hwa University, Hualien, Taiwan, Republic of China}
\affiliation{ $^2$ CERCA, Department of Physics, Case Western Reserve University, Cleveland OH 44106-7079}
\author{Dejan Stojkovic$^{3}$}
\affiliation{ $^3$ HEPCOS, Department of Physics, SUNY at Buffalo, Buffalo, NY 14260-1500, USA}

\begin{abstract}
\widetext
Using the complete classification of the bases in the rotating black hole background we separate superradiance from the Hawking effect.  We first find that there is spontaneous particle creation for fermions by the potential outside the black hole horizon for the frequencies inside the superradiant regime, i.e.  
$\omega<k\Omega_H$. However, these particles do not enhance the total flux from the black hole.  For the superradiance particle to became real, its negative energy counterpart has to be canceled by the positive energy Hawking radiation mode at the horizon. Since  due to the Pauli's principle this cancellation must be one-to-one, the superradiance effect cannot add anything to the total black hole flux.  For an extremal black hole, the Hawking temperature is zero, horizon is not populated with thermal modes, and fermions can be emitted through the superradiance mechanism.  On the other hand, a macroscopic flux of fermions infalling to the black hole is the opposite process of Hawking radiation. A positive energy-infalling particle must cancel out a negative energy thermal mode at the horizon, which leaves a net positive energy mode that crosses the horizon. Since there is finite thermal particle density at the horizon, this implies that there is a maximal fermion infalling rate which is also controlled by the Hawking temperature. 

\end{abstract}


\maketitle

\section{ Introduction}
The phenomenon of superrdiance, i.e. a process in which disordered energy is converted into coherent energy \cite{dicke}.
Most notably, in the presence of a potential barrier, superradiance is a classical phenomenon in which an amplitude of an outgoing wave after the reflection from the barrier is greater
than the amplitude of the ingoing wave \cite{Zeldovich}.  This effectively leads to a reflection coefficient greater than one (i.e. negative absorption coefficient). Superradiance can also happen in the background of a rotating black hole \cite{unruh,Unruh1973}, where an incident wave can take away part of the rotational energy of the black hole and get amplified after reflection. This has been described in a wide range of situations in the literature (see e.g.  \cite{Baryakhtar:2020gao,Rahmani:2020vvv,Khodadi:2020cht,Franzin:2021kvj,Chen:2021zqs,Richarte:2021fbi,Khodadi:2021owg,Ishii:2022lwc,Mascher:2022pku,Yang:2022uze,Jha:2022tdl,Alexander:2022avt,Cuadros-Melgar:2021sjy,Khodadi:2021mct,Khodadi:2022dyi,Brito:2015oca,Eskin:2015ssa,Vicente:2018mxl}).
 Nevertheless, it should be noted that the crucial difference between the Hawking effect and superradiance is that the Hawking effect happens in the presence of the horizon, while the superradiance does not need a horizon. Superradiant emission is simply the effect of particle creation in scattering from the potential barrier outside of the horizon. 
Some more recent papers raise a question whether horizon is really necessary even for the Hawking effect (e.g. \cite{Vachaspati:2006ki,Barcelo:2010pj,Wondrak:2023zdi}), suitable boundary conditions and the existence of some other dissipative mechanism can take the role of the horizon \cite{Richartz:2009mi,Cardoso:2012zn}. However, we adopt here the standard picture where the Hawking radiation is created when one member of a virtual pair created in vicinity of the horizon falls into the black hole while the other member escapes to infinity. 

In the context of black hole radiation, it has been noticed that  the superradiance is not possible for fermions \cite{Unruh1973,Frolov_book,CHANDRASEKHAR,Unruh:1974bw,Maeda:1976tm,Martellini:1977qf,Kim:1997hy,Lee:1977gk,Iyer:1978du}. It was found that a part of an incoming Dirac field gets reflected from the black hole horizon,  but in contrast with the bosonic fields,  the amplitude is never enhanced. The Pauli exclusion principle is often vaguely mentioned as a reason, however without any explicit description of the process.  In quantum field theory,  if a fermion flux is reflected by an electric field barrier, the amplitude of the reflected wave can be enhanced. Since the fermion flux can be enhanced in QED, but not in the background of a black hole, the black hole geometry must play a very important role. The aim of this paper is to clarify this issue using the complete classification of the bases in the rotating black hole background given in \cite{Frolov_book}. 

In this paper, we make a subtle distinction between the superradiance in the sense of spontaneous particle production by the gravitational potential in the superradiant regime of frequencies, and superradiance in the sense of amplification of the amplitude of the reflected wave. While the former is present for fermions in the background of rotating black holes, the latter is absent.

\section{ Equation of motion for fermions.}
\begin{figure}[h]
\includegraphics[width=8cm]{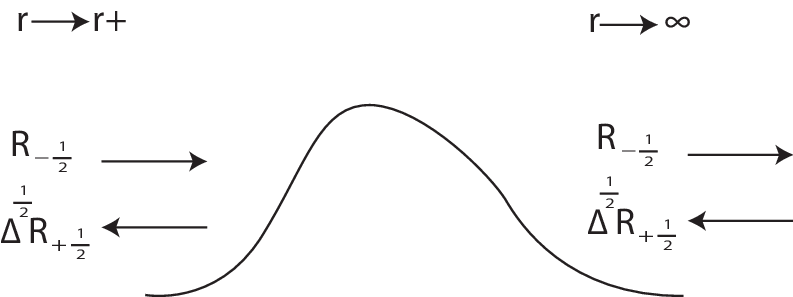}
\caption{ If $\omega>k\Omega_H $, then $R_{-\frac{1}{2}}$ is an out-going mode and  $\Delta^{\frac{1}{2}} R_{+\frac{1}{2}}$ is an in-coming mode at the horizon. If $\omega<k\Omega_H $, the roles are reversed, so  $R_{-\frac{1}{2}}$ is an in-coming mode and  $\Delta^{\frac{1}{2}} R_{+\frac{1}{2}}$ is an out-going mode at the horizon (not shown in the figure). The function  $Z_+$ field includes four possible asymptotic states at $r\rightarrow \infty$ and $r\rightarrow r_+$. A combination of these asymptotic states can represent a field coming from infinity which is  scattered by the potential. Part of it crosses the horizon and part of it is reflected back to infinity. It can also represent a wave which escapes from the horizon and is scattered by the potential. Part of it is transmitted through the barrier  and goes to infinity, and part of it is reflected and goes back to the horizon.   
}
\label{field}
\end{figure}

We start with the metric for a rotating black hole in the standard form
\begin{eqnarray}
ds^2 &=& -(1-\frac{2Mr}{\Sigma})dt^2 -\frac{4Mra\sin^2\theta}{\Sigma} dt d\phi +\frac{\Sigma}{\Delta}dr^2  \nonumber\\
&+&\Sigma d\theta^2+\Big(r^2+a^2 +\frac{2Mra^2\sin^2\theta}{\Sigma}\sin^2\theta \Big) d\phi^2,\\
\Delta &=& r^2 -2Mr +a^2,\\
\Sigma &=&r^2 +a^2\cos^2\theta ,
\end{eqnarray}
where $a$ is the black hole rotation parameter. The Newton's constant, $G$, Planck's constant, $\hbar$, the speed of light, $c$, and Boltzmann constant, $k_B$, are set to $1$.  The massless Dirac equation without an external potential is
\begin{equation}
(\gamma^\mu \nabla _\mu ) \Psi =0 , 
\end{equation}
where $\gamma^\mu$ are the general relativistic Dirac matrices, which satisfy
\begin{equation}
\{\gamma^\mu,\gamma^\nu \}=2g^{\mu\nu} .
\end{equation}
The metric connection is 
\begin{equation} 
\nabla_\mu =\partial_\mu +\frac{1}{8}\omega_{\mu\alpha\beta} [\gamma^\alpha,\gamma^\beta ] ,
\end{equation}
where $\omega_{\mu\alpha\beta}$ is the spin connection. The Dirac field, $\Psi$, is a 4-spinor, but it can be written in the chiral 2-spinor representation 
\begin{equation}
\Psi = \begin{bmatrix}
       P^A       \\[0.3em]
       \bar{Q}_{B'}   
     \end{bmatrix}
\end{equation}
where $P^A$ and $\bar{Q}_{B'}$ are the chiral eigenvectors. They represent the particle helicity (defining the left or right handed fermions) in the massless case.  Matrices $\gamma^\mu$ are 
\begin{equation}
\gamma^\mu =\sqrt{2}\begin{bmatrix}
       0_{C^2}  & \sigma^{\mu AB'}     \\[0.3em]
       \sigma^\mu _{AB'} & 0_{C^2}   ,
     \end{bmatrix}
\end{equation}
$P^A$ and $\bar{Q}_{B'}$ denote 2-component spinors, $\sigma^\mu_{AB'}$ are the Hermitian ($2\times 2$)-Infeld-van der Waerden symbols, and $A\in {1,2}$ and  $B'\in {1',2'}$. The 2-spinor form of the Dirac equation is 
\begin{eqnarray}
\sigma^\mu_{AB'}\nabla_\mu P^A  =0&\\
\sigma^\mu_{AB'}\nabla_\mu Q^A  =0 .
\end{eqnarray}
Here, the Pauli matrices are
\begin{equation}
\sigma^\mu _{(k)(l')} = \begin{bmatrix}
       l^\mu  & m^\mu     \\[0.3em]
       \bar{m}^\mu & n^\mu   ,
     \end{bmatrix}
\end{equation}
and the null  vectors are chosen to be 
\begin{eqnarray}
l^\mu&=&\frac{1}{\Delta}(r^2+a^2,\Delta,0,a)\\
n^\mu&=&\frac{1}{2\rho^2}(r^2+a^2,-\Delta,0,a)\\
m^\mu&=&\frac{1}{\bar{\rho} \sqrt{2}}(ia\sin\theta,0,1,i\csc\theta)\\
\bar{m}^\mu &=& m^{\mu*} .
\end{eqnarray}
Here, $\bar{\rho}=r+ia\cos\theta$. This 2-spinor Dirac equation can be separated into the radial and angular part by applying the following substitution
\begin{eqnarray}
P^0&=&\frac{e^{-i\omega t +ik\phi }}{\sqrt{2}(r-ia\cos\theta)}R_{-\frac{1}{2}}(r)S_{-\frac{1}{2}}(\theta)\\
P^1&=&e^{-i\omega t +ik\phi } R_{+\frac{1}{2}}(r)S_{+\frac{1}{2}}(\theta)\\
\bar{Q}_{0'}&=&e^{-i\omega t +ik\phi } R_{+\frac{1}{2}}(r)S_{-\frac{1}{2}}(\theta)\\
\bar{Q}_{1'}&=&\frac{e^{-i\omega t +ik\phi }}{\sqrt{2}(r+ia\cos\theta)}R_{-\frac{1}{2}}(r)S_{+\frac{1}{2}}(\theta) ,
\end{eqnarray}
where $k$ is the azimuthal number, $k\in Z+\frac{1}{2} $. Though it looks like $P^A$ and $\bar{Q}_{A'}$ are related, they just happen to satisfy similar relations, and must be considered independently. The Dirac equation now reduces to two pairs of equations
\begin{eqnarray}
\label{R1}
\Delta^{\frac{1}{2}}(\partial_r +\frac{iK}{\Delta}) R_{-\frac{1}{2}}&= & \lambda \Delta^{\frac{1}{2}}R_{+\frac{1}{2}}\\
\label{R2}
\Delta^{\frac{1}{2}}(\partial_r -\frac{iK}{\Delta}) \Delta ^{\frac{1}{2}}R_{+\frac{1}{2}}&= & \lambda R_{-\frac{1}{2}}\\
(\partial_\theta +Q+\frac{1}{2}\cot\theta)S_{+\frac{1}{2}}&=&-\lambda  S_{-\frac{1}{2}}\\
(\partial_\theta -Q+\frac{1}{2}\cot\theta) S_{-\frac{1}{2}}&=&\lambda  S_{+\frac{1}{2}}
\end{eqnarray}
Here, 
\begin{eqnarray}
K&=&-(r^2+a^2)\omega +ak\\
Q&=&-a\omega \sin\theta +k\csc\theta
\end{eqnarray}
 After redefining the radial coordinate as   
\begin{equation}
dr^*=\frac{r^2+a^2}{\Delta} dr ,
\end{equation}
the asymptotic solution can be found near the horizon and at infinity. 
When $r\rightarrow \infty$, we have
\begin{eqnarray}
R_{-\frac{1}{2}} &\sim& e^{i\omega r^*}\\
\Delta^{\frac{1}{2}} R_{+\frac{1}{2}} &\sim& e^{-i\omega r^*} .
\end{eqnarray}
Apparently, $R_{-\frac{1}{2}}$ represents an outgoing mode and  $\Delta R_{+\frac{1}{2}}$ an in-coming mode (Fig.~\ref{field}). 
When $r\rightarrow r^+$, we have
\begin{eqnarray}
R_{-\frac{1}{2}} &\sim& e^{i(\omega -k\Omega_H )r^*}\\
\Delta^{\frac{1}{2}} R_{+\frac{1}{2}} &\sim& e^{-i(\omega -k\Omega_H)r^*} ,
\end{eqnarray}
where $\Omega_H=\frac{a}{2Mr_+}$. For the modes with $\omega>k\Omega_H $, $R_{-\frac{1}{2}}$ is an out-going mode and  $\Delta^{\frac{1}{2}} R_{+\frac{1}{2}}$ is an in-coming mode near the horizon. For the modes with  $\omega<k\Omega_H $, $R_{-\frac{1}{2}}$ is an in-coming mode and  $\Delta^{\frac{1}{2}} R_{+\frac{1}{2}}$ is an out-going mode near the horizon (Fig.~\ref{field}). 

Following the procedure outlined in \cite{Chandrasekhar_book}, $R_{-\frac{1}{2}}$ and $\Delta^{\frac{1}{2}} R_{+\frac{1}{2}}$ are combined into a single wave function
\begin{equation}
Z_{+}=\Delta^{\frac{1}{2}} R_{+\frac{1}{2}}+ R_{-\frac{1}{2}} .
\end{equation}
The function  $Z_{+}$ represents a $1+1$-dimensional particle wave which interacts with the gravitational potential and then gets scattered away. 

\section{ Spontaneous particle creation outside the horizon.}
\begin{figure}[h]
\includegraphics[width=8cm]{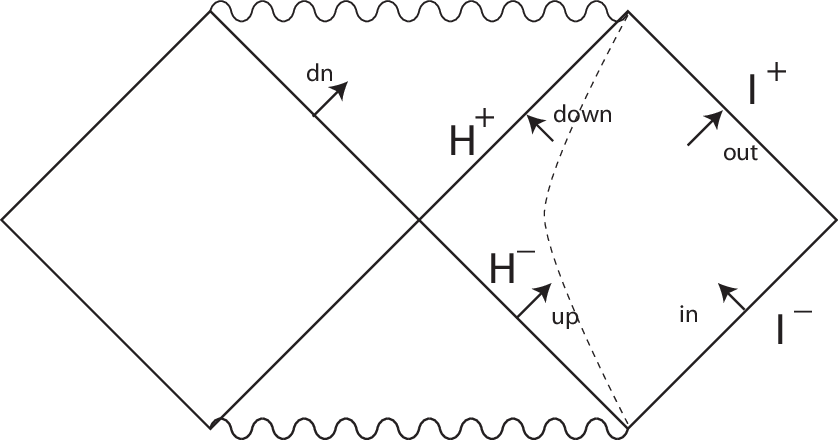}
\caption{The black hole Penrose-Carter diagram. The space outside the horizon is in the right square. The upper triangle represents the space inside the future horizon. The black hole radiation due to Hawking effect involves these two regions. On the other hand, only the right square is involved in particle creation by the superradiance mechanism. The thin dashed line represents the potential barrier that induces the superradiance. Four types of bases ($down$, $up$, $in$ and $out$) are involved in the process. Five bases (including $dn$) are involved in the full black hole radiation \cite{Frolov}. Similar argument for scalar field can be found in \cite{Dai:2023ewf}.        
}
\label{penrose-carter}
\end{figure}

The spontaneous particle creation related to superradiance happens in the region outside of the (past and future) horizon labeled by H$^+$ and $H^-$ in Fig.\ref{penrose-carter}. So all the events relevant for superradiance happen in the right square in Fig. \ref{penrose-carter}. In contrast, Hawking radiation is induced by the presence of the black hole horizon.

To study the process of spontaneous particle creation we have to identify the basis in which we decompose the fields, and the vacuum state of the field.
To describe a vacuum state of a field, we need at least two bases.  For our purpose, we define four possible bases \cite{Frolov}. The first is the $in$-coming mode, which represents a wave going from the past null infinity, $I^-$, to the black hole, 
\begin{equation}
\label{chi-1}
Z^{in}_{J}\sim\frac{1}{\sqrt{\omega}}\exp(-i\omega r_*) .
\end{equation}
The second one is the $out$-going mode, which represents a wave propagating from the black hole to the future null infinity, $I^+$, 
\begin{equation}
\label{chi-2}
Z^{out}_{J}\sim\frac{1}{\sqrt{\omega}}\exp(i\omega r_*)
\end{equation}
The third one is the $down$ mode, which represents a wave going into the future horizon, $H^+$, 
\begin{equation}
\label{chi-3}
Z^{down}_{J}\sim\frac{1}{\sqrt{|\omega - k\Omega_H|}}\exp(-i(\omega - k\Omega_H) r_*) .
\end{equation}
The forth one is the $up$ mode, which represents a wave going away from the past horizon, $H^-$, 
\begin{equation}
\label{chi-4}
Z^{up}_{J}\sim\frac{1}{\sqrt{|\omega - k\Omega_H|}}\exp(i(\omega - k\Omega_H) r_*) .
\end{equation}

Since a field decomposition requires four distinct bases, we can decompose $Z_+$ in two different ways  
\begin{eqnarray}
\hat{Z}_+ &=& \sum_J \hat{a}_J^{in} Z^{in}_J +  \hat{a}_J^{ up} \tilde{Z}^{up}_J +h.c.\\
&=&\sum_J \hat{b}_J^{out} Z^{out}_J +  \hat{b}_J^{ down} \tilde{Z}^{down}_J +h.c.
\end{eqnarray}
where $J=\{\omega,l,k,s\}$ with $s$ being the helicity of the fermion, while $h.c.$ stands for the hermitian conjugate terms. We also have
\begin{eqnarray}
 \tilde{Z}^{\alpha}_J&= &Z^{\alpha}_J\text{, if }\omega - k\Omega_H>0\\
 &= &Z^{\alpha}_J {}^*\text{, if }\omega - k\Omega_H<0 ,
\end{eqnarray}
where $\alpha$ can be $up$ or $down$. The two types of vacua corresponding to $\hat{a}_{J}^{\alpha}$ and  $\hat{b}_{J}^{\alpha}$ are 
\begin{eqnarray}
\hat{a}_{J}^{\alpha}\ket{in;0} &=& 0\\
\hat{b}_{J}^{\alpha}\ket{out;0} &=& 0 .
\end{eqnarray}

The $in$ mode is expressed in terms of the bases in the past, while the $out$ mode is expressed in terms of the bases in the future. Consider now a field which starts from the vacuum in the far past, and after evolving is seen in terms of the $out$ bases
\begin{eqnarray}
Z^{in}_J&\rightarrow& R_J Z^{out}_J+T_J Z^{down}_J \\
Z^{up}_J&\rightarrow&  t_J Z^{out}_J+r_J Z^{down}_J  .
\end{eqnarray}

The creation and annihilation operators are related according to the relationship between the modes. For $\omega - k\Omega_H>0$, we have
\begin{equation}
\hat{b}^{out}_J= R_J\hat{a}^{in}_J+t_J \hat{a}^{up}_J   .
\end{equation}
In this case, there is no particle creation since there is no mixing of the creation and annihilation operators. 
For $\omega - k\Omega_H<0$, we have
\begin{equation}
\hat{b}^{out}_J= R_J\hat{a}^{in}_J+t_J \hat{a}^{up}_J {}^\dagger .
\end{equation}
In this case there is particle creation because of the mixing of the creation and annihilation operators. Thus, $\omega - k\Omega_H<0$ is the necessary condition for the superradiance. 
From here we can calculate the particle creation number due to the superradiance effect as
\begin{equation}\label{sr}
n_J=\bra{in,0}b_J^{out} {}^\dagger b_J^{out}\ket{in,0}=|t_J|^2 \text{, if }\omega - k\Omega_H<0 .
\end{equation}
To compare the superradiance particle creation with the Hawking effect and demonstrate their difference, we calculate the total  particle creation number (Hawking effect plus superradiance) characterized by the transmission coefficient $|t_J|^2$ \cite{Hawking:1975vcx,Damour:1976jd}
\begin{equation} \label{tr}
n_J^T=\frac{|t_J|^2}{\exp(\frac{\omega - k\Omega_H}{T})+1}, 
\end{equation}
where, $T$ is the black hole temperature. 

From Eqs.~(\ref{sr}) and (\ref{tr}) we can see the fundamental difference  between the superradiance and Hawking effect. For example, for an extremal black hole we have $T\rightarrow 0$. If $\omega > k\Omega_H$, which is outside of the superradiant regime, $n_J^T=0$ since both the Hawking effect and superradiance are absent.  However, if $\omega < k\Omega_H$,  we get $n_J=n_J^T$. In that case the superradiance is present and it is the only contribution to the total radiation from a black hole \cite{Dai:2010xp}. This clearly indicates that there is a black hole spontaneous radiation for fermions produced by the potential barrier in addition to Hawking radiation.

\begin{figure}[h]
\includegraphics[width=8cm]{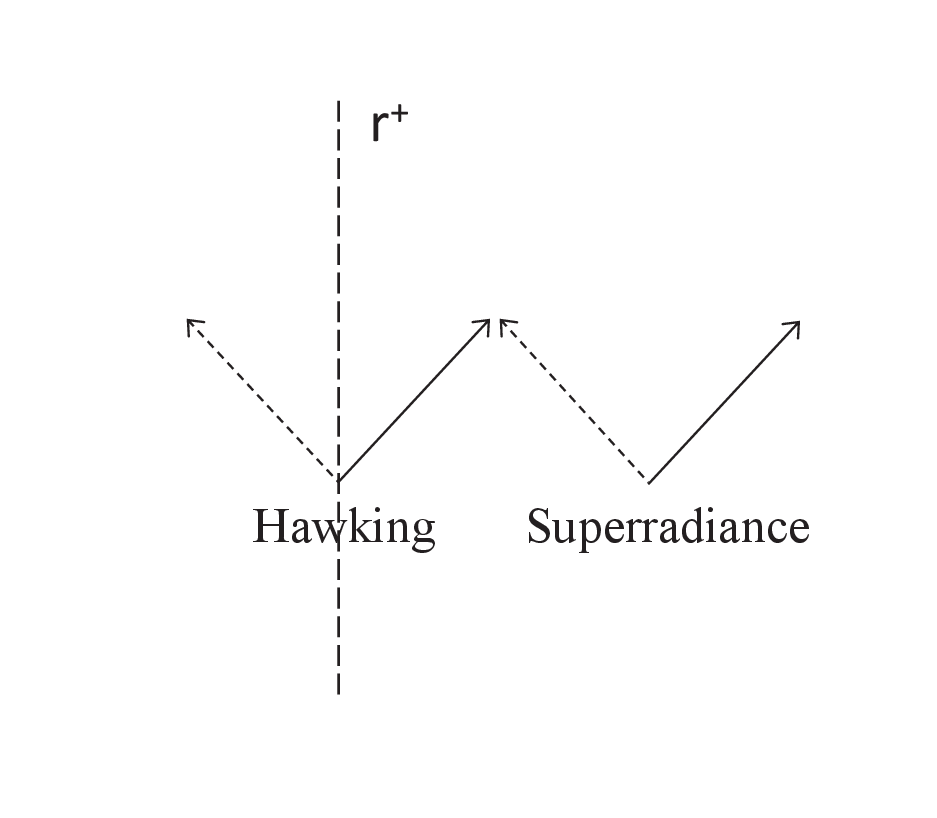}
\caption{Hawking effect happens at the horizon, while superradiance happens at the barrier outside of the horizon. For the superradiance particle to become real, its negative energy counterpart has to cancel out one of the thermal positive energy mode. The superradiance particle creation rate is higher than the black hole radiation rate. This means that the superradiance is suppressed by the presence of the horizon and cannot exceed the thermal particle production by the Hawking effect.              
}
\label{spontaneous}
\end{figure}

Important things happen for the superradiance modes $\omega < k\Omega_H$ when $T\neq 0$ in Eq.~(\ref{tr}).  The total black hole radiation, i.e. Eq.~(\ref{tr}), contains less modes than what  calculations without the horizon would give, i.e. Eq.~(\ref{sr}), since $\exp(\frac{\omega - k\Omega_H}{T})+1>1$. This implies that the existence of black hole suppresses the superradiance effect. This suppression is induced by the black hole's temperature, which also relies on particle creation from vacuum and reduces the number of the available negative energy modes. This of course happens because of the Pauli's exclusion principle which forbids two fermions to occupy the same state. More precisely, when a pair is created by the potential outside of the horizon (superradiance effect), for a positive energy particle to become real and leave the black hole, the negative energy particle must fall into the horizon. Hawking effect creates positive and negative energy modes at the horizon. So the negative energy mode from the superradiance cancels out the positive energy mode from the Hawking effect, as in Fig.~\ref{spontaneous}.  As the net effect, a positive energy particle leaves the black hole and negative energy particle gets absorbed by the black hole. The Hawking radiation modes and superradiance modes are traded one-to-one, so there is no net gain. Since we have more superradiance modes in the absence than in presence of the horizon, this means that the thermal black hole horizon suppresses superradiant vacuum fluctuations.  One might think that a negative superradiance mode could fall through the horizon directly, but it cannot since the Hawking effect is thermal, which means democratic, i.e. all the available modes at the horizon for that temperature are already occupied. So the  negative superradiance mode has to rely on the positive Hawking radiation mode to be absorbed.  It is not unexpected that the thermal Hawking flux suppresses out the superradiance modes since the high temperature moves particle distribution from low energy to high energy by reducing the low energy particle modes. And the superradiance modes are indeed the low energy modes.

\section{ Maximal infalling rate of fermions.}

This phenomenon has one more interesting consequence. If we send a flux of infalling (positive energy) fermions to the black hole, not all of them will be absorbed by a black hole. The black hole already has its saturated thermal fluctuations at the horizon, so only the infalling particles that can annihilate with the antiparticles at the horizon can contribute to the net flux through the horizon. If the infalling flux is beyond what the maximal thermal flux can support, the Pauli exclusion principle will apply again, which will in turn reduce the infalling rate, and possibly increase the reflection rate. This implies that there is a maximal infalling rate for fermions into a black hole  
\begin{equation}
\label{infall_rate}
F^J_{max}(\omega)\equiv  \frac{dN_{max}}{dt d\omega} = \frac{1}{e^{-\frac{\omega-k\Omega_H}{T}}+1} .
\end{equation}
$F^J_{max}$ is the maximal particle number per unit time and per unit energy, while $J$ stands for the quantum numbers of fermions.  
We note that we used the minus sign in the exponent of Eq.~(\ref{infall_rate}), i.e. we use the distribution of the negative energy modes. Because of the one-to-one trade, the maximum infalling rate depends on particle distributions both inside and outside of the horizon. On both sides, particles must satisfy the Pauli exclusion principle. We know that the horizon is thermalized, but the whole outer region (from the horizon to infinity) does not have to be thermalized. Therefore, we can only apply the constraint on particles at the horizon, so we only use the distribution of the negative energy particles as the suppression factor.   

If the flux is larger than this maximum value in Eq.~(\ref{infall_rate}), the excess of particles will not be absorbed by the black hole.  We note that this is a conservative estimate since we neglected the particles which are reflected back while trying to propagate from the horizon out through the barrier. This suppression is significant for high temperature black holes, so it could play an important role for small primordial black holes surrounded by fermions in plasma.  For large astrophysical black holes, low temperature implies more negative energy thermal particles at the horizon, so there are more negative energy modes that infalling particles can annihilate with, so the suppression is smaller. Also, for a fixed temperature, $T$, low energy particles, $\omega < k \Omega_H$ are suppressed more than high energy particles.

\section{ Conclusion.} To summarize, we separated the effect of superradiance from the total radiation from the black hole. We showed that the superradiance for fermions does exist in the sense of spontaneous particle production for the frequencies in the superradiant regime  by the potential barrier outside of the horizon. However, these particles do not enhance the reflection amplitude, nor the total flux from the black hole since every superradiance particle that became real was traded one to one with the thermal Hawking flux particle due to the Pauli's exclusion principle. 

The same mechanism imposes a maximal fermion infalling rate controlled by the black hole temperature. Smaller black holes with higher temperatures have more thermal modes the horizon, which in turn gives lower maximal infalling rate. One may wonder if the same conclusions can be drawn for a horizonless compact object. The answer is positive, if appropriate boundary conditions or dissipative mechanisms that take over the role of the horizon are applied. The negative energy flow at the horizon must be absorbed by some mechanisms, for example by particle pair annihilation or specific boundary conditions. In that case the maximum infalling rate should  be the same as in Eq.~(\ref{infall_rate}) with $T=0$.  
  
\begin{acknowledgments}
D.C. Dai is supported by the National Science and Technology Council (under grant no. 111-2112-M-259-016-MY3)
D.S. is partially supported by the US National Science Foundation, under Grant No.  PHY-2014021.  
\end{acknowledgments}

\end{document}